\documentclass[showpacs,twocolumn,amsmath,amssymb,prb,superscriptaddress]{revtex4}

\usepackage{graphicx}
\usepackage{amsmath,amsfonts,amssymb,amsthm,bm}
\usepackage{fancyhdr}
\usepackage{mathrsfs}
\usepackage{epstopdf}
\usepackage{color}

\begin{document}

\title{
      Antibunched photons from inelastic Cooper-pair tunneling 
}

\author{Juha Lepp\"akangas}
\affiliation{Microtechnology and Nanoscience, MC2, Chalmers
University of Technology, SE-412 96 G\"oteborg, Sweden}

\author{Mikael Fogelstr\"om}
\affiliation{Microtechnology and Nanoscience, MC2, Chalmers
University of Technology, SE-412 96 G\"oteborg, Sweden}

\author{Alexander Grimm}
\affiliation{Universit\'e Grenoble Alpes, INAC-SPSMS, F-38000 Grenoble, France}
\affiliation{CEA, INAC-SPSMS, F-38000 Grenoble, France}

\author{Max Hofheinz}
\affiliation{Universit\'e Grenoble Alpes, INAC-SPSMS, F-38000 Grenoble, France}
\affiliation{CEA, INAC-SPSMS, F-38000 Grenoble, France}

\author{Michael Marthaler}
\affiliation{Institut f\"ur Theoretische Festk\"orperphysik, Karlsruhe Institute of Technology, D-76128 Karlsruhe, Germany}

\author{G\"oran Johansson}
\affiliation{Microtechnology and Nanoscience, MC2, Chalmers
University of Technology, SE-412 96 G\"oteborg, Sweden}



\begin{abstract}
We demonstrate theoretically that charge transport across a Josephson junction, voltage-biased through a resistive environment,
produces antibunched photons.
We develop a continuous-mode description of the 
emitted radiation field in a semi-infinite transmission line
terminated by the Josephson junction.  Within a perturbative treatment
in powers of the tunneling coupling across the Josephson junction, we
capture effects originating in charging dynamics of consecutively
tunneling Cooper pairs.
We find that within a feasible experimental setup
the Coulomb blockade provided by high zero-frequency impedance
can be used to create antibunched photons at a very high rate
and in a very versatile frequency window ranging from a few GHz to a THz.
\end{abstract}

\maketitle



{\it Introduction}
Photons from ordinary thermal sources have a tendency to bunch
together, and the first controlled generation of single photons was
only performed in 1974~[\onlinecite{Clauser1974}]. Since then,
single photons have been used to explore fundamental aspects of
quantum mechanics, such as the interference fringes of single
particles~[\onlinecite{Grangier1986}] and Bell's
inequalities~[\onlinecite{Aspect1981}]. A good single-photon source
can be characterized by the vanishing probability of detecting two
photons at the same time from its output. This property is called
anti-bunching~[\onlinecite{Kimble1977}], and is also a sign that
the field is non-classical~[\onlinecite{WallsMilburn}].  In addition
to being of fundamental interest, the capability to create single
propagating photons is an indispensable tool for many quantum
information applications, including quantum key
distribution~[\onlinecite{Gisins}].

In the optical domain, the workhorse for many single-photon experiments has
been parametric down conversion~[\onlinecite{WallsMilburn}].  Here, photon
pairs are generated at random times, and one of the two photons can be
used to herald the other one~[\onlinecite{PDC}].  Other type of sources
for photons in the visible spectrum are based on
photoluminescence in systems such as single ions, single molecules,
semiconductor quantum dots and diamond color
centers~[\onlinecite{Review2005,Review2011}]. 

In the microwave domain, superconducting quantum
circuits~[\onlinecite{cQED1,cQED2,cQED3}] have been used to build
single photon sources
[\onlinecite{Houck2007,Bozyigit2001,Eichler2012,Lang2013,Yin2013,Pechal2014}]. In
most of these realizations, a two-level system is first excited and
consequently spontaneously emits a microwave photon into a
transmission line. The emission rate can often be controlled in-situ,
allowing for waveform shaping~[\onlinecite{Yin2013,Pechal2014,Pierre2014}]. 
The use of the field reflected from a single artificial
atom, which is indeed perfectly anti-bunched, has equally been
demonstrated~[\onlinecite{Hoi2012,Joel2014,Chang2007}]. In addition to fundamental quantum optics experiments, single photon
sources in the microwave domain might be useful for metrological purposes due to their well defined output power and, 
in combination with single photon detectors, for quantum nondemolition measurements
due to their ideal amplitude squeezing.

In this article, we study theoretically an alternative type of
single-photon source, based on the Coulomb blockade of charge transport across a Josephson junction in series with an 
electromagnetic environment~[\onlinecite{Fulton1989,SchonZaikin,Ingold1992}]. Our
proposed setup has a very simple operating principle and promises to
be a bright, robust and versatile 'photon gun'.

Anti-bunched photons are created from the applied voltage, making use
of inelastic (photon-assisted) Cooper-pair
tunnelling~[\onlinecite{Devoret1990,Ingold1992,Holst1994,Hofheinz2011}]
and the long charging time of the junction.  When biased below the
superconducting gap, the applied voltage defines the frequency
spectrum of the emitted photons through the Josephson
frequency, $\omega_{\rm J}/2\pi= 2eV/h$.   This frequency is not limited to the plasma frequency of the Josephson junction and the energy of the emitted photons is, in principle, only limited by the gap of the superconductor ($\hbar\omega_{\rm J}<4\Delta$).  Using appropriate materials it should thus be possible to reach frequencies from a few GHz to 1\,THz, making our source compatible with the energy scales of several other quantum systems including semiconductor quantum dots.
The high frequencies would also allow for single photon envelopes on the cm length scale,
facilitating experiments on quantum non-locality within a cryostat.

\begin{figure}[t]
\includegraphics[width=0.8\linewidth]{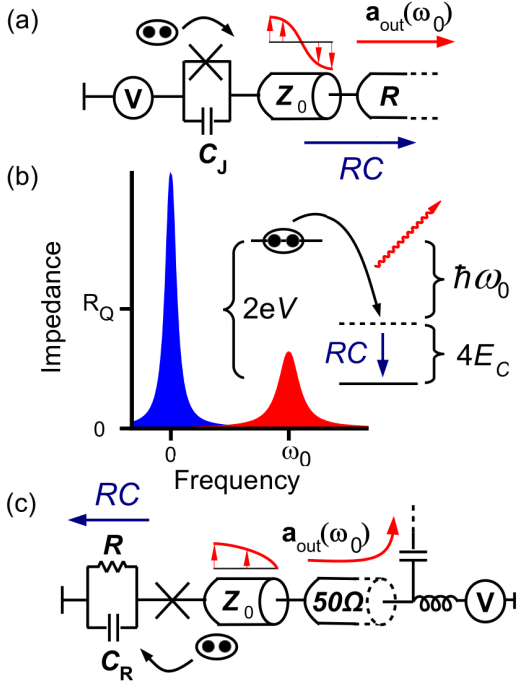}
\caption{(a)  The single-photon source we consider consists of a voltage-biased Josephson junction (JJ)
  and a parallel junction capacitor $C_{\rm J}$,
  in series with a semi-infinite transmission line (TL) with an impedance step, $Z_0<R$.
  The TL provides a $\lambda/2$-type standing wave between the step and the junction,
  and results in similar impedance as in (b-c).
(b) The desired impedance as seen by the JJ, presents a peak at zero
  frequency described by a zero frequency resistance $R  > h/4e^2=R_{\rm Q}$
  and a capacitance $C$ defining its bandwidth, as well a
  peak at finite frequency $\omega_0$ opening a window for photon
  emission. A photon can be emitted by Cooper-pair (CP) tunnelling when the voltage $V$ is chosen so
  that $2eV = 4e^2/2C + \hbar \omega_0$. Further CP tunnelling events are momentarily
  blocked because more energy would be needed to add a second CP to the effective capacitor $C$.
  After a time $RC$ the capacitor discharges
  and the next photon-assisted CP tunnelling can occur.
(c) An experimentally feasible realization allowing for large
  zero-frequency impedance $R\gg R_{\rm Q}$ while maintaining
  compatiblity with standard $50\,\Omega$ TLs. Here the voltage bias
  is applied to the junction via a $Z=50\,\Omega$ TL and a $\lambda/4$
  segment with $Z_0>50\,\Omega$. The same TL is used to collect the
  high-frequency response by splitting the signal in high- and low-frequency components.
  In this case the large resistance $R$ can be realized on chip as a thin-film resistor.
  }
\label{fig:Setup}
\end{figure}

The single-photon source we consider is shown in
figure~\ref{fig:Setup}. It consists of a Josephson junction biased at
voltage $V$ and embedded in an electromagnetic environment
characterized by the impedance $Z(\omega)$ as seen by the junction. The
impedance is engineered to be high at the desired frequency $\omega_0$,
opening a window for photon emission at this frequency,
as well as high at zero frequency ($Z(0) > h/4e^2=R_{\rm Q}$).
This 0-frequency peak is further characterized by a
capacitance $C$, in the simplest case the junction capacitance $C_{\rm J}$. When a
Cooper-pair tunnels across the junction, it gains the energy $2eV$
from the voltage source. However, the electrostatic energy of the
capacitor ($C$) also increases by the charging energy
$4E_C=2e^2/C$. For operation, the released energy is made to match the
frequency of the mode $\omega_0$, i.e. $2eV-4E_C=\hbar\omega_0$.  After
each photon-assisted tunnelling event, the high zero-frequency
impedance ($R > R_{\rm Q}$) and capacitance lead to a long
charging time $RC$ of the junction, which temporarily blocks further
tunnelling~[\onlinecite{Averin,Likharev,SchonZaikin,Pistolesi2012,LeppakangasBOT}]. This results in photon anti-bunching in the field emitted into the transmission
line. In this picture, the slow recharging dissipates the rest of the
energy supplied by the voltage source ($2eV-\hbar\omega_0=4E_C$)
in the form of a large number of emitted low frequency photons,
that in an experimental setup are absorbed by the physical resistor.

{\it Methods} To quantitatively test if such an interplay between the
low- and high-frequency parts of the same electromagnetic environment
is possible, we take a specific implementation of the environment
response function $Z(\omega)$, as described in figure~\ref{fig:Setup}(a),
and study the solution of the propagating continuous-mode flux field
$\hat \Phi(x,t)$ in the neighborhood of the Josephson junction ($x=0$) and in the
transmission line.  In the semi-infinite transmission line the
quantized field can be presented as a sum of incoming and outgoing
waves~[\onlinecite{WallquistPRB,Leppakangas2013,Leppakangas2014}],
\begin{eqnarray}\label{eq:WaveGeneral}
&&\hat \Phi(x,t)=\sqrt{\frac{\hbar R}{4\pi}}\int_0^\infty\frac{d\omega}{\sqrt{\omega}}\times \\
&\times&\left[ \hat a_{\rm  in} (\omega)e^{-i(k_\omega x+\omega t)}+\hat a_{\rm out} (\omega)e^{-i(-k_\omega x+\omega t)}+{\rm H.c.} \right].  \nonumber
\end{eqnarray}
Here, $R=\sqrt{L'/C'}$ is the characteristic impedance, expressed via inductance $L'$ and capacitance $C'$ per unit length,
and $k_{\omega}=\omega\sqrt{C'L'}$ is the wave number.
The incoming (outgoing) wave at frequency $\omega$ is created by the operator $\hat a_{\rm in}^{\dagger}(\omega)$ ($\hat a_{\rm out}^{\dagger}(\omega)$)
and annihilated by $\hat a_{\rm in}(\omega)$ ($\hat a_{\rm out}(\omega)$).
They fulfill the standard commutation relation $\left[ \hat a_{\rm in}(\omega),\hat a_{\rm in}^{\dagger}(\omega')  \right]=\delta(\omega-\omega')$.
A similar solution exists also in the cavity region (impedance $Z_0 < R$).

The interaction between the radiation field and Cooper-pair tunnelling across the Josephson junction
is described by the boundary condition ($x=0$ corresponds to junction location)
\begin{eqnarray}\label{eq:BoundaryCondition}
C_{\rm J}\ddot{\hat \Phi}(0,t)-\frac{1}{L_0'}\frac{\partial\hat\Phi(x,t)}{\partial x}\vert_{x=0}=I_{\rm c}\sin\left[ \omega_{\rm J} t - \hat \phi(t) \right]. 
\end{eqnarray}
Here $C_{\rm J}$ is the junction capacitance, $L_0'$ the inductance  per unit length in the region between the junction and the impedance step.
The Josephson current is limited by the critical current $I_{\rm c}$ and controlled by the phase $\hat \phi(t)\equiv 2\pi\hat \Phi(0,t)/\Phi_0$. 
Equation~(\ref{eq:BoundaryCondition})
corresponds to current conservation at the junction.
The boundary condition at the impedance step is linear and can be solved by Fourier transformation~[\onlinecite{Leppakangas2014}].
These two conditions can now be used to solve the free-space out-field $\hat a_{\rm out}(\omega)$ as a function of the equilibrium in-field
$\hat a_{\rm in}(\omega)$. The general solution for the out-field can be written formally as
\begin{eqnarray}\label{eq:OutputSolution}
\hat a_{\rm out}(\omega)&=& r(\omega)\hat a_{\rm in}(\omega) + i I_{\rm c} A(\omega)\sqrt{ \frac{Z_0}{\hbar\omega\pi}} \int_{-\infty}^{\infty}dte^{i\omega t} \nonumber \\
&\times &  \hat U^{\dagger}(t,-\infty)  \sin\left[\omega_{\rm J}t - \hat \phi_0(t) \right] \hat U(t,-\infty).
\end{eqnarray}
Here $A(\omega)$~(see Supplemental Material) is related to the impedance as seen by the Josephson junction,
${\rm Re}[Z(\omega)]\equiv Z_0\vert  A(\omega)\vert^2$,
and $r(\omega)= A(\omega)/A(\omega)^*$ corresponds to the phase shift in the out-field when $I_{\rm c}=0$.
The solution is expressed via time-evolution (operator) of the current across the Josephson junction~[\onlinecite{Ingold1998,Ingold1999,Leppakangas2015}],
\begin{equation}\label{eq:TimeEvolution}
\hat U(t,t_0)={\cal T } \exp\left\{ \frac{i}{\hbar}\int_{t_0}^{t} dt' E_{\rm J}\cos\left[\omega_{\rm J}t' - \hat \phi_0(t') \right] \right\}.
\end{equation}
Here $E_{\rm J}=(\hbar/2e)I_{\rm c}$ is the Josephson coupling and $\cal T$ stands for time-ordering.
This is an expansion in terms of the phase at the Josephson junction in the absence of the tunnelling current,
\begin{eqnarray}\label{PhaseDifference0}
\hat\phi_0(t)=\frac{\sqrt{4\pi\hbar Z_0}}{\Phi_0}\int_{0}^{\infty}  \frac{d\omega}{\sqrt{\omega}}  A(\omega)\hat  a_{\rm in}(\omega)e^{-i\omega t}+{\rm H.c.}
\end{eqnarray}

From here on, we make the natural assumption that temperature is low compared to
the mode frequency, $k_{\rm B}T\ll \hbar\omega_0$. 
The flux density of photons due to Cooper-pair tunnelling can be evaluated to
the leading order in the critical current $I_{\rm c}$~[\onlinecite{Hofheinz2011,Leppakangas2013,Leppakangas2014}],
\begin{eqnarray}\label{eq:OutputFlux}
f(\omega)&=& \int_0^{\infty} d\omega '\frac{1}{2\pi}\left \langle \hat a^{\dagger}_{\rm out}(\omega)\hat a^{}_{\rm out}(\omega') \right\rangle\\
&=&\frac{I_{\rm c}^2  {\rm Re}[Z(\omega)] }{2\omega} \ P(2eV-\hbar\omega). \nonumber
\end{eqnarray}
The photon flux is a function of the well-known probability density~[\onlinecite{Ingold1992}],
\begin{equation}\label{eq:PEFunction}
P(E)=\int_{-\infty}^{\infty} dt \frac{1}{2\pi\hbar} e^{J(t)}e^{i\frac{E}{\hbar}t},
\end{equation}
where $J(t)=\left\langle [\hat\phi_0(t)-\hat\phi_0(0)]\hat\phi_0(0) \right\rangle $ is a measure of equilibrium phase fluctuations
and is a function of input impedance and temperature~[\onlinecite{Ingold1992}].
The function $P(E)$ describes the ability of the transmission line to absorb an energy $E$ when a Cooper-pair tunnels.
In the setup we consider, the $P(E)$ function has a simple analytical form 
\begin{eqnarray}\label{eq:PEHighOhmic}
&&P(E) \approx (1-p) P_{ \rm CB }(E)+p P_{  \rm CB }(E-\hbar\omega_0),
\end{eqnarray}
where $p\ll 1$ (Supplemental Material) and $P_{  \rm CB }(E)$ is the probability distribution for a high-Ohmic
impedance, $R=Z_0\gg R_{\rm Q}$ with cut-off $1/RC$,
\begin{equation}
P_{  \rm CB }(E)=\frac{1}{\sigma\sqrt{2\pi}} e^{-\frac{1}{2\sigma^2}(E-4E_C)^2}.\nonumber
\end{equation}
Here, the broadening $\sigma=\sqrt{8E_C k_{\rm B}T}$ (assuming $1/RC<k_{\rm B}T/\hbar$)
describes thermal fluctuations of the junction voltage
induced by the series resistor.
These fluctuations are accounted for as interaction with (thermal) photons in the transmission line, treated to all orders.
The total $P(E)$-function has not only a peak at $E=4E_C$, but also at $E=4E_C+\hbar\omega_0$,
that corresponds to the possibility to emit the energy $\hbar\omega_0$ to the lowest cavity mode and
to use the rest to charge of the junction capacitor. 

\begin{figure}[tb]
\includegraphics[width=\linewidth]{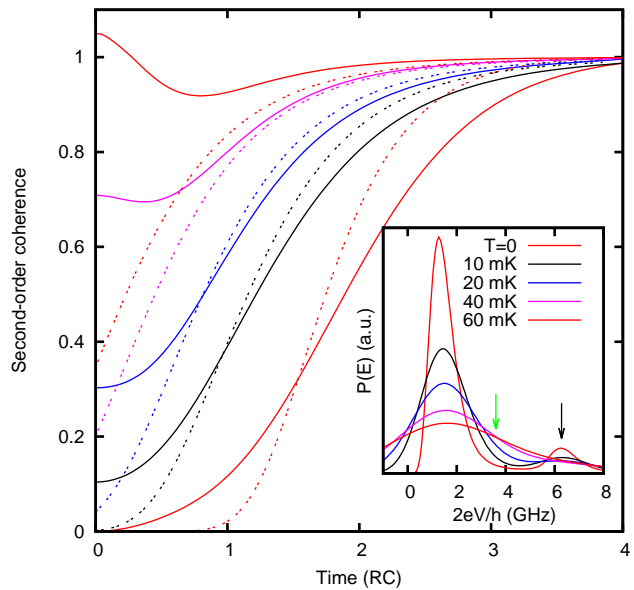}
\caption{
Second-order coherence $g^{(2)}(\tau)$ based on equations~(\ref{eq:ResultG2}-\ref{eq:BunchingMainResult}) (solid lines)
and on semi-analytical approximation~(\ref{eq:G2HighOhmic}) (dashed lines), at different temperatures.
When biased optimally (the black arrow) we observe anti-bunching within $RC$ time-scale and limited by temperature.
The expected voltage immediately after the first tunnelling is pointed by the green arrow.
The long-time behavior is reproduced by the semi-analytical formula, based on calculating the leading-order emission flux with a
slowly evolving voltage.
At short times, thermal fluctuations increase the bunching and wash out anti-bunching approximately when $T=60~{\rm mK}\approx 3E_C/k_{\rm B}$.
(Other parameters are given in the Supplemental Material.)
}
\label{fig:resultsT}
\end{figure}

{\it Results} We now investigate the probability to observe two photons with a time separation $\tau$. This is usually quantified in terms of the second order coherence function $g^{(2)}(\tau)$~[\onlinecite{Loudon}]. To calculate this, we need the first order coherence function, defined
via the photon flux as
\begin{equation}
G^{(1)}(\tau)=  \frac{R\hbar}{2}\int_0^{\infty} d\omega e^{i\omega\tau} \omega \vert F(\omega)\vert^2  f(\omega). \nonumber
\end{equation}
Here, $F(\omega)$ describes the measurement band-pass filter~[\onlinecite{Miranowicz2010}], centered at the chosen measurement frequency, $\omega \sim \omega_{\rm J}-4E_C/\hbar$.
The inverse of its bandwidth $W$ determines the temporal resolution of the photon detection.
The expression for the unnormalized second-order correlation function of photon detection reads
\begin{eqnarray}\label{eq:SecondCoherence}
&&G^{(2)}(\tau)\equiv\\
&& \left( \frac{\hbar R}{4\pi} \right)^2 \int_{\rm BW} e^{i\tau(\omega_2-\omega_3)} \sqrt{\omega_1\omega_2\omega_3\omega_4} \left\langle\hat a_{\omega_1}^{\dagger}\hat  a_{\omega_2}^{\dagger}\hat  a_{\omega_3} \hat a_{\omega_4} \right\rangle,\nonumber
\end{eqnarray}
%
%
using the shorthand notation 
$\int_{\rm BW}\equiv \Pi_i\int_{0}^{\infty}d\omega_i F(\omega_i)$ and $a_{\omega_i}\equiv a_{\rm out}(\omega_i)$.
The dominating contribution comes from the fourth order in $I_{\rm c}$ (Supplemental Material),
\begin{eqnarray}\label{eq:ResultG2}
&&G^{(2)}_{\rm 4th}(\tau)= \left(\frac{I_{\rm c}^2 R}{4\pi\hbar}\right)^2 \int_{\rm BW}  e^{i\tau(\omega_2-\omega_3)} A_{\omega}  \\
&& \int_{\rm times} \left \langle {\cal T}^{\dagger}\left\{ \hat I^{\dagger}_{\omega_1,t_1} \hat I^{\dagger}_{\omega_2,t_2} \right\}  \  {\cal T}\left\{\hat I_{\omega_3,t_3} \hat I_{\omega_4,t_4}\right\} \right \rangle,  \nonumber
\end{eqnarray}
where  $A_\omega\equiv A^*(\omega_1) A^*(\omega_2) A(\omega_3)A(\omega_4)\delta(\omega_1+\omega_2-\omega_3-\omega_4)$
and $\int_{\rm times}\equiv\Pi_i\int_{-\infty}^{\infty}dt_i$.
The operator $\hat I_{\omega,t}= e^{i\omega t} \exp\left[ i [ \hat \phi_0(t)- \omega_{\rm J} t]\right]$  can be interpreted
as a creation of a photon with frequency $\omega$ via Cooper-pair tunnelling
at time $t$. 
The final expression for the second order coherence can be presented as
\begin{equation}\label{eq:BunchingMainResult}
g^{(2)}(\tau)\equiv \frac{ G^{(2)}_{\rm 4th}(\tau) }{\vert G^{(1)}_{\rm 2nd}(0)\vert^2} = 1 + \left[ \frac{G^{(1)}_{\rm 2nd}(\tau)}{G^{(1)}_{\rm 2nd}(0)} \right]^2 + {\cal G}(\tau),
\end{equation}
where ${\cal G}(\tau)$ captures correlations between consecutive Cooper-pair tunnelling events
and the other terms describe the emission from uncorrelated Cooper-pair tunnelling. When the temporal resolution of the detection cannot resolve individual tunnel events, i.e.~for $W \rightarrow 0$, we get ${\cal G}(\tau) \rightarrow 0$ and  $g^{(2)}(0) \rightarrow 2$, which is characteristic for chaotic (thermal) light, having no phase coherence between different photons~[\onlinecite{Loudon}]. For times much longer than the temporal resolution $\tau \gg W^{-1}$ and recovery time $\tau \gg RC$, only the first term is finite leaving $g^{(2)}(\tau)=1$, a characteristic of uncorrelated (Poissonian)
 photon emission.
Outside these limits, we see clear effects of ${\cal G}(\tau)$, as shown in Fig.~\ref{fig:resultsT}, where we plot the second-order coherence calculated numerically using equations~(\ref{eq:ResultG2}-\ref{eq:BunchingMainResult}).
We consider a cavity length such that a resonance mode appears at $\omega_0/2\pi = 5$~GHz,
for $R=4R_{\rm Q}$ and $Z_0/R=1/10$. The detection filter is centered at the mode frequency with a bandwidth of $1$~GHz.
The $P(E)$-function (inset) is close to the analytical form in Eq.~(\ref{eq:PEHighOhmic}) and the simple interpretation
presented in figure~{\ref{fig:Setup}} implies clear anti-bunching [$g^{(2)}(0)<1$] of the outgoing radiation.
Indeed, when biased optimally at $2eV=4E_C+\hbar\omega_0$ (black arrow in the inset),
we observe a clear anti-bunching  over  a time-scale given by the $RC$-time.
The depth of the anti-bunching dip is limited by temperature, and increasing it beyond
the charging energy, $k_{\rm B}T \gtrsim 3E_C$, the output radiation instead becomes bunched [$g^{(2)}(0)>1$].


\begin{figure}[bt]
\includegraphics[width=\linewidth]{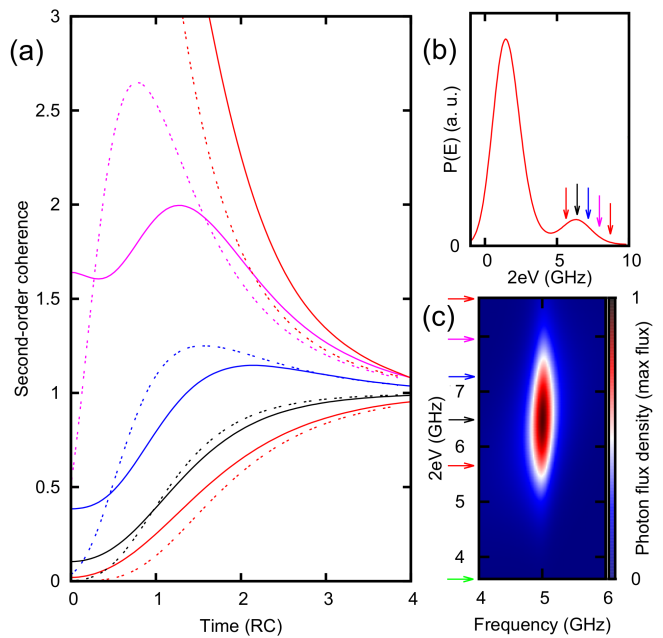}
\caption{
(a) Second-order coherence $g^{(2)}(\tau)$ for different bias voltages, based on equations~(\ref{eq:ResultG2}-\ref{eq:BunchingMainResult}) (solid lines) and on approximation~(\ref{eq:G2HighOhmic}) (dashed lines), for the system as in figure~\ref{fig:resultsT} with $T=10$~mK.
(b) The $P(E)$-function and the positions of the studied voltage-bias points, marked by the arrows of the corresponding colors.
(c) The photon-flux density as a function of the bias voltage in the neighborhood of the studied mode at $\omega_0/2\pi = 5$~GHz.
Depending on the bias point, the second-order coherence can have a deep antibunching dip,
a local maximum, or show a steep decay, as a function of the measurement separation-time $\tau$.
The steep decay or local maximum occurs since above the optimal bias (black arrow)
the first emission rate is decreased,
while the rate for the secondary emission can be always higher, or sweep through the maximum, during the recharging.
}
\label{fig:resultsG2}
\end{figure}

Classically, at zero temperature, a sudden $2e$-reduction of the equilibrium capacitor charge ($Q=CV$)
induces recharging dynamics $v(t)=V-(2e/C)e^{-t/RC}$. 
A first guess would be that (in the full quantum treatment) the probability to detect a second photon after a time $\tau$
is given by the photon emission rate at voltage $v(\tau)$.
Following this simple idea, we write down a semi-analytical formula for the expected second order coherence,
\begin{equation}\label{eq:G2HighOhmic}
g^{(2)}(\tau) = \frac{\int_{\rm BW} {\rm Re}\left[Z(\omega)\right] P\left[2e v(\tau)-\hbar\omega \right]}{\int_{\rm BW} {\rm Re}\left[Z(\omega)\right] P(2eV-\hbar\omega)}.
\end{equation}
In figure~\ref{fig:resultsT}, we see that equation~(\ref{eq:G2HighOhmic}) reproduces the long-time behavior, whereas
can fail when the separation between the tunnelling events becomes short,
i.e.~when the tunnelling processes start to overlap.
This can also be understood more mathematically, as detailed in the Supplemental Material.
The form implies that to suppress $g^{(2)}(0)$ at the optimal bias point,
we need the width of the $P(E)$ to be smaller than the bias jump $4E_C$, which leads to the conditions $k_{\rm B}T \ll E_C$ and
 $R\gg R_{\rm Q}$.

For further illustration of recharging effects in this system, we study the dependence of bunching on the bias voltage, see figure~\ref{fig:resultsG2}.
For high bias voltages the second-order coherence shows super-bunching ($g^{(2)}(0)> 2$),
whereas for lower bias voltages both anti-bunching and bunching can co-exists for a single curve.
These properties are qualitatively reproduced by equation~(\ref{eq:G2HighOhmic}),
and occurs here since the first emission rate is decreased compared to the optimal point ($2eV=4E_C+\hbar\omega_0$),
while the rate for the secondary emission is always higher, or sweeps through the maximum, during the recharging.
We also observe that $g^{(2)}(0)$ can be suppressed below the value of the optimal point,
by a bias-voltage $2eV<4E_C+\hbar\omega_0$, with the cost of a reduced photon flux.
This is a result of a reduction in thermal-fluctuation triggered two-photon emission processes.

In conclusion, by extending the well-known $P(E)$-theory to two-Cooper-pair processes
(i.e.\ to 4th order in $E_{\rm J}$) we have shown
that the Coulomb blockade of Cooper-pair tunneling 
allows for creation of strongly anti-bunched microwave photons from a simple dc bias under
experimentally realistic conditions.
The validity of our theoretical approach is limited to
tunnelling rates lower than $1/RC$, but we expect anti-bunching to survive also at higher tunnelling rates.
This regime would be characterized by modified junction voltage dynamics,
and is indeed an interesting regime to explore, both experimentally and theoretically. 
The proposed source of antibunched photons can be transformed into a photon gun
by replacing the Josephson junction by a dc-SQUID.
Cooper-pair tunnelling can then be efficiently suppressed by threading a
half flux quantum through the SQUID and a flux pulse short compared to
$RC$ will then release a single photon on demand with high probability
for large $E_{\rm J}$. Therefore, this system allows for an on-demand photon emission
at a very high rate and in a wide range of frequencies.
Such a bright microwave single photon source would be useful in various
microwave quantum measurement setups because of its perfect amplitude
squeezing and well defined power.

JL, MF and GJ acknowledge financial support from the Swedish Research Council and the European Union represented by the EU STREP project PROMISCE.
AG and MH acknowledge financial support from the Grenoble Nanosciences
Foundation and from the European Research Council under the European
Union's Seventh Framework Programme (FP7/2007-2013) / ERC Grant
agreement No 278203 -- WiQOJo.

\pagebreak
\widetext
\begin{center}
\textbf{\large Supplemental Material}
\end{center}
\setcounter{equation}{0}
\setcounter{figure}{0}
\setcounter{table}{0}
\setcounter{page}{1}
\makeatletter
\renewcommand{\theequation}{S\arabic{equation}}
\renewcommand{\thefigure}{S\arabic{figure}}
\renewcommand{\bibnumfmt}[1]{[S#1]}
\renewcommand{\citenumfont}[1]{S#1}


\subsection*{Factor $A(\omega)$}
Fourier transformation of the boundary conditions at the Josephson junction and at the impedance step leads to
\begin{equation}
A(\omega)=\frac{2\sqrt{\frac{Z_0}{R}} e^{-2ik^0_{\omega}d}}{\left(1+\frac{Z_0}{R} \right)e^{-2ik_\omega^{0} d}{\cal C}(\omega)+\left(\frac{Z_0}{R}-1\right){\cal C}^{*}(\omega)}.
\end{equation}
Here ${\cal C}(\omega)=1-iZ_0C_{\rm J}\omega$,
$k^0_{\omega}=\omega\sqrt{C_0'L'_0}$ is the wave number, and  $d$ is the length of the cavity.
A detailed mathematical derivation of $A(\omega)$ is given in reference~[35] (where the corresponding function is called $\bar A(\omega)$).

\subsection*{$P(E)$-function}
The real part of the impedance seen by the tunnel junction can be deduced to be ${\rm Re}[Z(\omega)]=Z_0\vert A(\omega)\vert^2$ (see reference~[35]).
In the case $Z_0\gg R$ the transmission line has ($\lambda/4$-type) resonance frequencies approximately at
$\omega_n=(2n+1)\omega_{\rm r}$, where $\omega_{\rm r}=\pi/2 d\sqrt{C_0'L_0'}$ and $n\in[0,1,2,\ldots ]$.
Near each resonance  the real part of the tunneling impedance is approximately a Lorentzian.
In the neighbourhood of the lowest mode $\omega_0$ we can write (keeping the notation similar with reference~[25]),
\begin{equation}
{\rm Re}[Z(\omega)]\approx \frac{1}{C_{\rm t}}\frac{\Gamma}{1+4(\omega-\omega_0)^2\Gamma^2}\approx \frac{\pi}{2 C_{\rm t}}\delta(\omega-\omega_0).
\end{equation}
Here $C_{\rm t}=\pi/4\omega_0 Z_0$ and $\Gamma=\pi Z_0/4R\omega_0$. 
The corresponding $P(E)$-function, in the $\delta$-function approximation and at zero temperature, has the form~[25]
\begin{equation}
P(E)=e^{-p}\sum_{k=0}^{\infty} \frac{p^k}{k!}\delta(E-k\hbar\omega_0).
\end{equation}
Here $p=(4e^2/2C_{\rm t})/\hbar\omega_0$ compares the effective capacitive energy of a single Cooper pair to the mode frequency.
In the case $Z_0\gg R$ we have $p=4Z_0/R_{\rm Q}$.
For a finite linewidth of the mode the resulting $P(E)$-function broadens accordingly.

In the case considered in this article, $Z_0\ll R$,
the transmission line has ($\lambda/2$-type) resonance frequencies approximately at
$n\omega_{\rm r}$, where $\omega_{\rm r}=\pi/ d\sqrt{C_0'L_0'}$ and $n\in[0,1,2,\ldots ]$.
We have therefore a peak also at zero frequency.
We can take here the approximation ${\cal C}(\omega)\approx 1$ which leads to
\begin{equation}
{\rm Re}[Z(\omega)]\approx \frac{R}{1+\left( \pi\frac{R}{Z_0}\frac{\omega}{\omega_0} \right)^2}.
\end{equation}
This means that the cut-off frequency of the zero-frequency peak is $1/RC$, where $C=\pi/Z_0\omega_0$.
For the used parameter range this effective capacitance dominates the junction capacitance ($C \approx 2C_{\rm J}$).

The effect of the junction capacitance $C_{\rm J}$ is more pronounced for the first resonance where it induces a shift to
a lower frequency and a reduction of the impedance.
We find an approximative correction to the resonance frequency in the
considered parameter region,
\begin{equation}
\omega_0\approx \omega_{\rm r}\frac{\pi+\frac{Z_0\omega_{\rm r}}{R\omega_{\rm c}} }{ \pi+2\frac{Z_0\omega_{\rm r}}{R\omega_{\rm c}}  }.
\end{equation}
Here we use the notation $\omega_{\rm c}=1/RC_{\rm J}$. In the considered case we have $\omega_{\rm 0}/\omega_r\approx 5/7$.
This allows for an analytical expression for the impedance in the neighbourhood of $\omega_0$.
However, the form of the Lorentzian in this case is rather cumbersome and less useful.
According to our numerical calculations, the corresponding probability parameter satisfies $p\ll 1$.
This can also be deduced from the relative heights of the first two peaks in the $P(E)$-functions plotted in figures 2-3,
bearing in mind equation~(S3).
This observation, and that the total $P(E)$-function of a sum of two tunnel impedances
is a convolution of the two $P(E)$-functions obtained for individual impedances, leads to equation~(8).

\subsection*{Second-order coherence function $G^{(2)}$}
Two-photon emission is characterized by the second-order coherence,
\begin{eqnarray}
G^{(2)}(\tau)\equiv \left( \frac{\hbar R}{4\pi} \right)^2 \int_{\rm BW} e^{i\tau(\omega'-\omega'')} \sqrt{\omega\omega'\omega''\omega'''} \left\langle\hat a_\omega^{\dagger}\hat  a_{\omega'}^{\dagger}\hat  a_{\omega''} \hat a_{\omega'''} \right\rangle . \nonumber
\end{eqnarray}
An expansion to the leading-order (second-order) in the junction's critical current describes photon emission from single Cooper-pair tunneling.
The straightforward calculation is presented in reference~[35], and involves the summation,
\begin{equation}\label{eq:GeneralCorrelator}
\left\langle \hat a^{\dagger}\hat a^{\dagger}\hat a\hat a \right\rangle \rightarrow \sum_{i,j,k,l} \left\langle \hat a_i^{\dagger}\hat a_j^{\dagger}\hat a_k\hat a_l \right\rangle,
\end{equation}
done with the constriction $i+j+k+l=2$. The sub-indices correspond to the order (in the critical current) of operators $\hat a$.
When $i=0$ or $l=0$, we have a contribution proportional to the Bose factor at the measurement frequency, 
which can be neglected in the frequency range we are interested in, $\hbar\omega\gg k_{\rm B}T$.
The main contribution comes when $i=l=1$ (and $j=k=0$), giving the unnormalized second-order coherence,
\begin{eqnarray}\label{eq:BunchingLeadingOrder} 
G^{(2)}_{\rm 2nd}(\tau) \propto  \int_{\rm BW} e^{i\tau(\omega'-\omega'')} A_\omega \  P[\hbar(\omega_{\rm J}-\omega-\omega')]  . \nonumber
\end{eqnarray}
This is characterized by the probability $P(\hbar\omega_{\rm J}-\hbar\omega-\hbar\omega')$.
This means that in the leading order the observed photons originate from single CP tunneling processes,
each of them releasing an energy $\hbar\omega_{\rm J}$, and that this energy has to be split into two individual photon energies.
At the frequency range we are interested in, $\omega\sim \omega_0$, such
photon pair production is negligible as long as $k_{\rm B}T/\hbar \ll \omega_0$.
Therefore, it is the fourth-order contribution in the critical current [equation~(10)] that
is dominating at low temperatures. A derivation of this contribution is given in reference~[39].

\subsection*{Semi-analytical formula}
For an analytical expression of the second-order coherence,
we consider the limit of a wide bandwidth (high temporal resolution), and that the field operators
\begin{equation}
\hat a(\tau)\sim\int_{\rm BW} e^{i\omega (t+\tau)} A(\omega) \hat I_{t,\omega},
\end{equation}
can be treated local in time (discussed in more detail below). This means that in the following $t_1<t_2$ and $t_3>t_4$.

We then study contraction of outer and inner tunneling operators in equation~(10).
We use the property of bosonic pair correlations,
\begin{eqnarray}
\left\langle\Pi_i \exp\left[n_i\hat \phi_0(t_i)\right] \right\rangle = \exp\left[ -\sum_{i<j}n_in_j J(t_i-t_j)  \right],\nonumber
\end{eqnarray}
where $n_i$ is an integer 1 or -1, and $\sum_i n_i=0$. We get,
\begin{eqnarray}
G_{\rm 4th}^{(2)}(\tau) &\propto& \int_{\rm BW} \int_{\rm times} e^{-i\omega_1 t_1}e^{-i\omega_2 (t_2+\tau)}e^{i\omega_3 (t_3+\tau)}e^{i\omega_4 t_4} \nonumber \\
& \times &A_\omega  e^{ J(t_1-t_4)+J(t_2-t_3)}e^{J_{\rm PI}}.
\end{eqnarray}
Here we have introduced a function that describes interaction between the contraction pairs,
\begin{equation}
J_{\rm PI}= J(t_1-t_3)+J(t_2-t_4)-J(t_1-t_2)-J(t_3-t_4).
\end{equation}
If $ J_{\rm PI} = 0 $, the pairs decouple
and the result of the time integration is proportional to a product of two $P(E)$-functions,
$P(\hbar\omega_{\rm J}-\hbar\omega)\times P(\hbar\omega_{\rm J}-\hbar\omega')$.
Here $\omega=\omega_1=\omega_4$ and $\omega'=\omega_2=\omega_3$.
This decoupling occurs always for long $\tau$ (which means here long $t_2-t_1$ or equivalently long $t_3-t_4$).
It can also occur for intermediate $\tau$ with a modified energy argument, as demonstrated below.

Let us consider the case $J_{\rm PI}\neq 0$ and when the most relevant contribution to the second-order coherence comes from short-time behaviour inside the pairs, $\vert t\vert=\vert t_1-t_4\vert <\tau$ and $\vert t'\vert=\vert t_2-t_3\vert <\tau$,
i.e.~there is a fast decay of the term $e^{ J(t_1-t_4)+J(t_2-t_3)+F}$ with increasing $t$ or $t'$.
This helps us because the time-ordering between the pairs is fixed and the integrations between the pairs do not overlap
(as assumed at the beginning, $t_1<t_2$ and $t_3>t_4$).
When $k_{\rm B}T/\hbar< 1/RC$, the time-scale of the decay is of the order, or less than, $RC$.
This means that we are safe (at low temperatures) if $\tau>RC$.


In the next step, when estimating the term $J_{\rm PI}$, we notice that all the time arguments of the individual phase
correlation functions are of the order $\tau$. This allows us to
consider an approximation of the phase-correlation function (when evaluating $J_{\rm PI}$),
\begin{equation}
J(t)=c -D' \vert t\vert-i \bar V't,
\end{equation}
where $D',\bar V'$, and $c$ are constants for a fixed $\tau$. This is a linear approximation for times in the neighbourhood of $\tau$.
The time-dependent imaginary part reads $\bar V'(\tau)\propto V'(\tau)=(2e/C)(1-e^{-\tau/RC})$,
corresponding to a momentary value during a classical RC-recovery of the voltage after a Cooper pair tunneling event.
Inserting approximation~(S10) to equation~(S9) leads to $J_{\rm PI}=-2i\bar V'(t_2-t_3)$, i.e.~$J_{\rm PI}$ is purely imaginary.
This term adds to the imaginary part of $J(t_2-t_3)$ in equation~(S8) and describes a shift in the junction voltage
$V\rightarrow V-V'$. Integration over the times and frequencies then lead to approximation~(12),
\begin{equation} 
g^{(2)}(\tau) = \frac{\int_{\rm BW} {\rm Re}\left[Z(\omega)\right] P\left[2e V'(\tau)-\hbar\omega \right]}{\int_{\rm BW} {\rm Re}\left[Z(\omega)\right] P(2eV-\hbar\omega)}.
\end{equation}

In the case of a high temperature, $k_{\rm B}T/\hbar \gg 1/RC$, the phase-correlation function $J(t)$
at intermediate times is quadratic rather than linear. This leads to the failure of the formula
in a wider range of time $\tau$ as for $k_{\rm B}T/\hbar \ll 1/RC$.
The general failure at short times can also be understood as a neglection of correlated voltage fluctuations
between the Cooper-pair tunnelings, and
as a neglection of two-photon emission via $4e$-tunnelling ("virtual" initial Cooper-pair tunnelling).
The formula also neglects interference effects between photons that lack relative phase memory,
which increases $g^{(2)}(0)$.

\subsection*{Parameters in the numerical simulations}
In figures 2 and 3 we use a free-space transmission line $R=4R_{\rm Q}$ and an impedance step $Z_0/R=1/10$.
Here $R_{\rm Q}=h/4e^2\approx 6.5~k\Omega$ is the superconducting resistance quantum.
The resonance frequency ($\omega_0/2\pi$) is at $5$~GHz and
the junction capacitance is $C_{\rm J}=25$~fF.
The measurement band-pass filter $F(\omega)$ was chosen to be a Gaussian with a full width at half maximum $1$~GHz.
As discussed above, the effective capacitance $C$ differs from the junction capacitance,
and becomes $C=50$~fF. The value is determined from the RC recovery of the imaginary part of $J(t)$.
We have therefore a charging energy $E_C=1.6$~$\mu$eV ($k_{\rm B}\times 18.6$~mK) and a charging time $RC=1.3$~ns.
Equations (10)-(11) were evaluated numerically using a three dimensional fast Fourier transformation.
Due to a finite discretization of the time and frequency spaces ($2^{3N}$ points, where $N=9$),
the plotted values of $g^{(2)}(\tau)$ can have a small numerical error,
which we estimate to be limited by $\pm 0.02$.

\subsection*{Feasibility of the voltage bias and high resistance environment}
Regarding the experimental feasibility, one question is the level of control of the bias voltage and temperature.
Thermal fluctuations in the theory of figure~1(a) are accounted for by the model as it treats
the interaction with bosonic modes in the bias line to all orders.
The experimentally relevant setup, figure 1(c), behaves in the same way,
but here the effective temperature has two contributions;
one from the on-chip resistor $R$, and one from fluctuations in the applied voltage.
Using low-impedance cold voltage dividers, the fluctuations of the applied bias voltage can be made as low as a few 100~nV (see reference [28]),
which is small compared to the voltage fluctuations of the on-chip resistor at 10~mK (several~$\mu$V).
Therefore the relevant difficulty is the cooling of the on-chip resistor.

The on-chip resistance $R > R_{\rm Q}$ can be realized, for example, using a thin-film chromium wire.
Thermalisation of on-chip resistors is rather well established and is discussed in detail in Phys.~Rev.~B {\bf 76}, 165426 (2007).
We explore in figure 2 how cold the resistor must be for our proposal to work and obtain experimentally achievable temperatures.
The resistor will have stray capacitances, but in the acceptable range for the proposal to work.

\subsection*{Feasibility of the on-demand photon emission}
In the last part of the article we describe how one can realize a photon gun by using a SQUID rather than one Josephson junction.
By changing the magnetic  flux through the SQUID one can reach the right regime of photon emission for a controlled amount of time.
One concern here could be that hysteresis in the junction dynamics would hinder the relaxation to the desired voltage bias condition.
Hysteresis is avoided since the flux pulse is used to tune $E_{\rm J}$ to zero,
so that there will be no nonlinear effect due to the junction which could lead to hysteresis.
If this flux value is maintained for a time long compared to the relaxation-time of the electromagnetic environment,
that is here $RC$, any sub-gap voltage can be reached.
The increase (and decrease) of $E_{\rm J}$
should be fast compared to RC but slow compared to $1/\omega_0$, not to create photons by the pulse itself. 
If $E_{\rm J}$ is increased slowly compared to $RC$ instead, the junction voltage can latch to 0 after a photon has been emitted,
due to the possibility of consecutive Cooper-pair tunnelings with no photon emission.
This effect can actually be beneficial for our proposal, as it deepens the anti-bunching (but goes beyond the theory developed in this paper).
We also note that sub-gap quasiparticle tunneling will not be induced,
when the applied flux-pulse is slow in the timescale of the (inverse) superconducting gap.

\end{document}